\documentclass[9pt, conference]{IEEEtran}
\IEEEoverridecommandlockouts

\usepackage{cite}
\usepackage{amsmath,amssymb,amsfonts}
\usepackage{algorithmic}
\usepackage{graphicx}
\usepackage{textcomp}
\usepackage{xcolor}
\usepackage{authblk}
\def\BibTeX{{\rm B\kern-.05em{\sc i\kern-.025em b}\kern-.08em
    T\kern-.1667em\lower.7ex\hbox{E}\kern-.125emX}}

\usepackage{booktabs}
\usepackage{array}
\usepackage{multirow}
\usepackage{url}
 
\newcommand{\dbits}{\Delta\text{bits}}
\newcommand{\dauc}{\Delta\text{AUC}}
\newcommand{\dfa}{\Delta\text{FA}}

\begin{document}

\title{How Much Prosody Does Your ASR Encoder Already Know}

\author{Rini Sharon$^{\star}$ \qquad Kadri Hacioglu$^{\star}$ \qquad Andreas Stolcke$^{\star}$}

\affil{$^{\star}$ Uniphore} 



\maketitle


\begin{abstract}
End-of-turn (EoT) detection decides in real-time whether a speaker has finished. Traditional
detectors rely on hand-built prosodic features (pitch, speaking rate, pauses), while modern
detectors run on a neural ASR encoder that has processed the same audio. Because the encoder reads
the signal those features are computed from, it may already hold much of the information they were
meant to supply. We test this three ways, reading a streaming Zipformer encoder at four depths from raw input
to the final layer. First, we measure how much prosody helps over the encoder alone. Second, we
test whether prosody can be recovered from the encoder. Third, we ask whether any single feature
still helps given the encoder. The three tests agree; prosody adds a large gain on raw audio, but
that gain collapses once the signal passes through even the first encoder layer. The same pattern holds for two open-source encoders, wav2vec2 and Whisper, both of which are offline.
\end{abstract}

\begin{IEEEkeywords}
end-of-turn detection, prosody, streaming speech recognition, turn-taking,
representation analysis
\end{IEEEkeywords}

\section{Introduction}
\label{sec:intro}

A spoken dialogue system must work out when the user has finished speaking before it can respond,
and it must make this end-of-turn (EoT) decision continuously and at low latency. The timing is
demanding because human turn-taking is fast. In conversation the gap between speakers is usually
only a few hundred milliseconds \cite{stivers2009,levinson2015}, so a mistimed response is noticed
at once. Responding too early cuts the user off, while responding too late makes the exchange feel
sluggish \cite{skantze2021review}. The simplest approach is to wait for a fixed
stretch of silence and treat it as the end of the turn. Silence is easy to measure, but on its own
it is unreliable, because speakers also pause in the middle of a turn. A fixed silence threshold
then has to choose between two failures, firing on these within-turn pauses or waiting too long
after the turn has ended \cite{heldner2010,raux2008}.

To do better than silence alone, end-of-turn detection has long relied on prosody, the cues in
\emph{how} something is said rather than \emph{what} is said, such as pitch, energy, speaking
rate, and pause structure \cite{gravano2011,skantze2021review}. Speakers mark the end of a turn
with these cues, for instance by lowering their pitch and slowing down, a regularity documented
across decades of work in conversation analysis and phonetics
\cite{fordthompson1996,sacks1974,duncan1972}. Early detectors used the cues directly. Combining
pitch, vowel duration, and speaking rate, for example, detects the end of an utterance faster and
more accurately than a fixed silence threshold \cite{ferrer2002,ferrer2003}, and dialogue systems
used similar timing cues to decide when to stop listening and respond \cite{raux2008}. Prosody
thus became a turn-end signal in its own right, computed by hand straight from the audio, and it
is still studied as a turn-taking cue in recent work \cite{ekstedt2022prosody,inoue2024}.

Modern detectors are built differently, yet they share one assumption, that prosody adds something
the encoder lacks. They sit on top of a neural \emph{encoder}, a network trained for automatic
speech recognition (ASR) that turns the raw audio into the sequence of feature vectors a
recogniser consumes \cite{gulati2020,zipformer2024,radford2023whisper}. Across two decades of such
systems, prosody and the encoder's features have stayed separate inputs, combined by design. Early
neural detectors replaced the fixed silence rule with learned acoustic models that pick up cues
such as filler sounds and speaking rate \cite{shannon2017}, and streaming variants made the
decision online, updating the turn-end estimate at each new frame \cite{masumura2017online}. Later
systems built endpointing into the recogniser itself, so that one model both transcribes and marks
turn ends \cite{maas2018,chang2019}, while others predicted from moment to moment whether the user
is about to keep talking or stop \cite{skantze2017,roddy2018investigating,ekstedt2022vap}. The
practice continues today, as current turn-taking models still pair a dedicated acoustic or prosodic
stream with the learned representation, whether they predict turns from voice activity
\cite{inoue2024} or build on a large language model \cite{wang2024turntaking}, and a recent survey
still singles out prosody as a central turn-taking cue \cite{skantze2021review}. At the same time,
the newest full-duplex speech LLMs fold turn-taking directly into a single end-to-end model with no
separate prosodic front-end \cite{defossez2024moshi,veluri2024syncllm}, already hinting that turn
dynamics can live inside a learned representation.

The encoder, though, reads the same audio the prosodic features are computed from, so it may
already represent much of that prosody on its own \cite{zipformer2024,gulati2020}. Properties such
as pitch and stress can be recovered from a speech encoder's internal representations by probing, a
standard analysis technique with established safeguards against reading in correlations that are
not really there \cite{pasad2021,delafuente2024,belinkov2022,hewitt2019,voita2020}. A
self-supervised turn-taking model has also been shown to change its predictions when the prosody of
its input audio is altered \cite{ekstedt2022prosody}, and a recent study probes whether such
turn-taking models rely on prosodic or lexical cues, finding either can suffice \cite{russell2026}.
None of this answers the practical question.
With a strong encoder already in the pipeline, does adding hand-built prosody still improve
detection, or has the encoder made it redundant? If it is redundant, the field has been paying for
feature extraction, latency, and per-speaker tuning for little benefit.

This paper answers the question directly, reading a streaming Zipformer ASR encoder at four depths, from the
raw input features through to the final layer, and running three tests.
\begin{itemize}
  \item \textbf{Does prosody still help?} We train a turn-end detector on the encoder alone and
  again on the encoder together with prosody, and measure how much prosody adds at each depth
  (Sec.~\ref{sec:res-curve}).
  \item \textbf{If not, why not?} We measure how accurately each prosodic feature can be
  \emph{reconstructed} from the encoder's own representations. A feature that is easy to reconstruct
  yet adds nothing when supplied separately is one the encoder already carries, which would explain
  why adding it again does not help (Sec.~\ref{sec:res-recon}).
  \item \textbf{Does any individual cue still help?} We test whether any single prosodic feature
  carries turn-timing information the encoder lacks, including the pitch cues classical detectors
  relied on most (Sec.~\ref{sec:res-subset}).
\end{itemize}

A small measured gain could mislead us for two reasons, and we control for both. The first is that
the detector might be too weak to use the prosody it is given, which would make prosody look
useless even when it carries real information. We rule this out by repeating every measurement with
a second, stronger detector and checking whether the conclusion holds. The second is that part of a
feature's apparent value can come from who the speaker is rather than from how the turn unfolds.
Take a speaker's mean pitch, which barely changes within a turn, so on its own it says little about
when that turn ends, yet it can still help a detector if speakers with higher voices happen to take
turns differently from speakers with lower voices. The detector then leans on the speaker's
identity as a shortcut instead of the turn-timing cue we care about
\cite{geirhos2020,lapuschkin2019,torralba2011}. We isolate and remove this speaker effect with a
\emph{time-shift control} that circularly shifts each prosodic feature off the turn boundary while
leaving its per-speaker statistics intact (Sec.~\ref{sec:method-detectors}), so that each feature
is credited only for genuine turn-timing information. We also repeat the main
measurement on two further open-source encoders and on a public corpus, to test whether the pattern is
specific to a single model or to our own data (Sec.~\ref{sec:res-generality}).
Section~\ref{sec:method} sets out each measurement, control, and data split in full.

\section{Dataset and Methodology}
\label{sec:method}

Every measurement in this paper rests on one shared setup, namely a single dataset, two fixed
detectors, and a time-shift control that separates genuinely turn-aligned timing from
speaker-identity effects. We define that setup here and present the experiments and their results
in Sec.~\ref{sec:results}.

\subsection{Task and labels}
\label{sec:method-task}
We work at the frame level, reading each utterance as a sequence of frames at $25$\,Hz, one every
$40$\,ms, and at each frame the detector predicts whether the turn ends there. The ground truth is
a single reference turn-end time $t^\ast$ per turn, the labelled instant at which the speaker
actually finishes.

These reference times come from our annotation pipeline. Each call is
two-party telephone speech with one speaker per channel, and $t^\ast$ marks the moment the speaker
yields the floor. In a two-party call this is every turn boundary where the floor passes to the
other speaker, together with the final turn of a call, where the speaker stops and no one takes
over. Each cut is extended past $t^\ast$ by up to $1.3$\,s of trailing silence, checked to contain
little speech, and labelled at $100$\,Hz, with frames before $t^\ast$ set to $0$ and frames from
$t^\ast$ onward through the extension set to $1$.

\begin{figure}[t]
\centering
\includegraphics[width=\columnwidth]{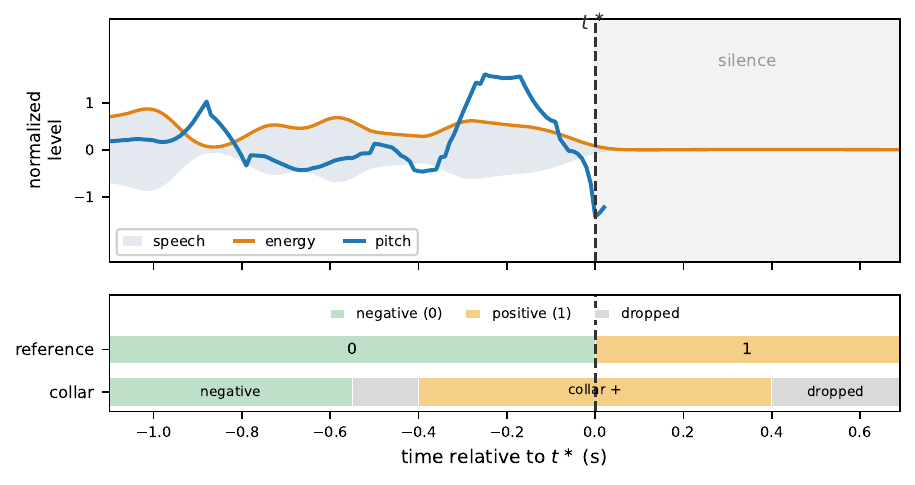}
\caption{A turn end from the data. The top panel shows the waveform with normalised energy and
pitch contours falling into the boundary $t^\ast$ before silence. The bottom panel shows the native
reference label ($0$ before $t^\ast$, $1$ through the silence extension) against our collar label,
whose positives sit in a window around $t^\ast$ while the guard band and trailing silence are
dropped.}
\label{fig:pipeline}
\end{figure}

Two frame rates are in play, since these reference labels are produced at $100$\,Hz ($10$\,ms),
the rate of the annotation pipeline, whereas the encoder and detector operate at $25$\,Hz
($40$\,ms), a $4\times$ downsampling. We map the $100$\,Hz labels onto the $25$\,Hz grid the
detector reads and state every window below in $25$\,Hz frames. We do not train on the native
$0/1$ labelling directly, because the trailing silence is labelled entirely positive and no
silence segment is labelled negative, so a detector can score well simply by firing on silence
without locating the transition at $t^\ast$. We therefore score on a \emph{collar} around
$t^\ast$, where frames within $W$ of $t^\ast$ are positive, frames well before it are negative, and
a short guard band together with the trailing silence are dropped from training and evaluation
alike (Fig.~\ref{fig:pipeline}). We use $W{=}10$ frames ($\pm400$\,ms) as our main setting and also
report $W\in\{5,15\}$. Because the collar keeps only the frames around the boundary, where
turn-final prosodic cues are concentrated \cite{gravano2011,fordthompson1996}, a finding that
prosody is redundant under this label is conservative.

\subsection{Data and representations}
\label{sec:method-feats}
We evaluate on a fixed sample of $2{,}221$ conversational cuts from $1{,}354$ speakers, with
$100$\,Hz reference EoT labels, drawn from an internal evaluation set of English conversational
telephone speech that spans $16$ application domains (distinct client call types, or verticals).
The sample is balanced across these $16$ domains, drawing a comparable number of cuts from each so
that no single domain dominates the reported numbers, and splits are speaker-disjoint throughout
(Sec.~\ref{sec:method-ci}).

The prosodic side is a bank of $41$ frame-level features spanning five families, each grounded in
an established turn-taking or endpointing cue (Table~\ref{tab:feats}), computed at $100$\,Hz and
pooled to $25$\,Hz by taking the mean and maximum in each window.

\begin{table}[t]
\caption{The $41$ prosodic features by cue family, with the turn-taking work motivating
each as a turn-end cue. Spectral-shape descriptors are carried for completeness, with no
specific turn-taking prior.}
\label{tab:feats}
\centering
\scriptsize
\setlength{\tabcolsep}{3pt}
\renewcommand{\arraystretch}{1.1}
\begin{tabular}{@{}>{\raggedright\arraybackslash}p{0.16\columnwidth} >{\raggedright\arraybackslash}p{0.78\columnwidth}@{}}
\toprule
family & features (motivating work) \\
\midrule
pitch / F0 &
\texttt{f0\_norm}, \texttt{f0\_delta}, \texttt{f0\_range}, \texttt{f0\_slope},
\texttt{f0\_mean\_local}, \texttt{f0\_std\_local}, \texttt{pitch\_contour\_type},
\texttt{pitch\_acceleration}, \texttt{f0\_slope\_300ms\_abs},
\texttt{f0\_slope\_signed\_500ms}, \texttt{f0\_slope\_at\_voicing\_offset},
\texttt{f0\_range\_1s}, \texttt{pitch\_reset}
~\cite{duncan1972,fordthompson1996,gravano2011,bogels2015,wennerstrom2003,delooze2014} \\
\addlinespace
energy &
\texttt{rms\_energy\_norm}, \texttt{energy\_delta}, \texttt{energy\_slope},
\texttt{energy\_drop}~\cite{gravano2011,local1986} \\
\addlinespace
timing / pause &
\texttt{speaking\_rate\_local}, \texttt{syllable\_rate}, \texttt{speech\_burst\_duration},
\texttt{pause\_after\_speech}, \texttt{time\_since\_last\_pause},
\texttt{silence\_duration}, \texttt{silence\_duration\_norm}, \texttt{speech\_flag}
~\cite{wightman1992,heldner2010,ferrer2002,ferrer2003,raux2008,shannon2017} \\
\addlinespace
voice quality &
\texttt{voicing\_strength}, \texttt{voicing\_onset\_rate\_1s}, \texttt{hnr\_ratio},
\texttt{jitter}~\cite{ogden2001,gravano2011,skantze2021review} \\
\addlinespace
spectral &
\texttt{spectral\_tilt}, \texttt{spectral\_rolloff}, \texttt{spectral\_bandwidth},
\texttt{spectral\_flatness}, \texttt{zero\_crossing\_rate},
\texttt{spectral\_contrast\_0}--\texttt{3}, \texttt{mfcc\_delta\_1}--\texttt{3}
~\cite{skantze2021review} \\
\bottomrule
\end{tabular}
\vspace{-0.4cm}
\end{table}

The encoder side is a streaming Zipformer ASR encoder \cite{zipformer2024}, read at four depths:
the input log-mel features ($80$-dim), the output of the first encoder stack (zip-early, $192$-dim),
the third stack (zip-mid, $512$-dim), and the final encoder output (zip-final, $768$-dim). The
encoder has six stacks in all, which we also read individually for a finer depth sweep
(Sec.~\ref{sec:res-curve}). We extract these activations from a single non-streaming forward pass,
so that forward hooks fire on every stack, and verified that its final-layer output matches the
deployed streaming causal encoder almost exactly (cosine $1.0$ on cuts shorter than the encoder's
left context and $\geq 0.98$ on longer cuts). We therefore analyse offline activations that
faithfully represent the streaming encoder.

\subsection{Metrics}
\label{sec:method-metrics}
Every comparison is an \emph{increment}, in that we never score prosody in isolation but ask how
much a detector improves when prosody is added to a representation it already has. Let $R$ denote
the encoder features and $P$ the prosodic features, and let $[R,P]$ be their concatenation. We
train the same detector on $R$ and on $[R,P]$ and summarise the gain from $P$ with two paired
measures of
end-of-turn detection,
\begin{equation}
\begin{aligned}
\dbits &= \mathrm{CE}(y\mid R)-\mathrm{CE}(y\mid R,P),\\
\dauc &= \mathrm{AUC}(y\mid R,P)-\mathrm{AUC}(y\mid R),
\end{aligned}
\label{eq:dbits}
\end{equation}
where $y$ is the frame's turn-end label and the conditioning set $R,P$ denotes the concatenation
$[R,P]$ of the two feature sets, not their sum. $\mathrm{AUC}(y\mid\cdot)$ is the area under the ROC
curve, the probability that a randomly chosen turn-end frame scores above a randomly chosen
non-turn-end frame, so $\dauc$ measures how much prosody improves the detector's ranking of frames.
$\mathrm{CE}(y\mid\cdot)$ is the held-out cross-entropy in bits per frame, so $\dbits$ measures how
much prosody sharpens the detector's predicted probabilities. Both are defined so that $0$ means
prosody adds nothing once $R$ is present and a positive value means it helps.

$\dbits$ has a clean information-theoretic reading. The held-out cross-entropy $\mathrm{CE}(\cdot)$
is the average number of bits needed to encode the true turn-end labels using the detector's
predicted probabilities, and it is small when the detector predicts those labels well and large
when it predicts them poorly. $\dbits$ is the reduction in this coding cost when prosody is added to
the encoder features. Under a minimum-description-length view \cite{voita2020}, that reduction is
exactly the turn-end information prosody carries about the labels beyond what the encoder already
supplies, so a larger $\dbits$ means prosody contributes more information the encoder lacked. The
quantity is detector-relative, in that it measures the information a given detector can exploit
rather than a bound on the information actually present, so a stronger detector may extract more
from the same features and move $\dbits$ in either direction. We therefore rerun every measurement
with a second detector (Sec.~\ref{sec:method-detectors}) and read the effect through the time-shift
control rather than trusting any single $\dbits$ value.

AUC captures ranking performance, the probability that a turn-end frame outscores a non-turn-end
frame, and we report it because it is the standard detection metric. But once the encoder alone
already ranks frames well its AUC sits near its ceiling, so any gain prosody adds is small and hard
to read from $\dauc$ alone. We therefore lead with $\dbits$, which still registers such gains, and
report $\dauc$ alongside it.

Our second experiment (Sec.~\ref{sec:res-recon}) asks how much of each prosodic feature the encoder
already represents, by trying to predict the feature from the encoder's own activations, since a
feature the encoder reconstructs well is one it already carries. We score this reconstruction with
the held-out coefficient of determination,
\begin{equation}
R^2 = 1 - \frac{\sum_i (y_i - \hat{y}_i)^2}{\sum_i (y_i - \bar{y})^2},
\label{eq:r2}
\end{equation}
where $y_i$ is the true feature value, $\hat{y}_i$ the prediction, and $\bar{y}$ the training-split
mean. An $R^2$ of $1$ is perfect reconstruction and $0$ is no better than always predicting the
training mean.

\subsection{Detectors and controls}
\label{sec:method-detectors}
We use the same detector throughout, held fixed across conditions so that any difference in a metric
reflects the representation and not the detector. Two detectors play this role, and we name them
here and refer to them by name in the experiments.

Our main detector is the \emph{logistic-regression detector}, an $L_2$-regularised logistic
regression solved with L-BFGS \cite{pedregosa2011}. We standardise every input dimension using
training-split statistics, which puts features of different scales on a common footing and lets the
$L_2$ penalty act evenly across them. Turn-end frames are rare, so we fit on a class-balanced
subsample with equal numbers of turn-end and non-turn-end frames, and the detector therefore cannot
score well simply by predicting the majority class. We tune the regularisation strength separately
for each representation, by a grid search over $C\in[0.01,10]$ that maximises held-out AUC under
four-fold cross-validation. The folds are speaker-grouped, keeping all frames from a speaker
together, so tuning is never scored on a speaker it has already seen. Tuning per representation
gives each $R$ its own best-fit penalty, so the paired $R$-versus-$[R,P]$ comparison reflects the
value of $P$ rather than a difference in how well $R$ happened to be regularised.

As a capacity check, a \emph{small neural network} repeats every measurement. It uses a single
hidden layer of $64$ units, regularised with $L_2$ weight decay and early stopping. We recalibrate
its outputs with Platt scaling so that its predicted probabilities, and with them the bits metric,
stay meaningful \cite{guo2017}. This network is the stronger of the two detectors, so if an effect
appears only with it, the logistic detector was too limited to use the information and we trust the
neural answer.

The \emph{time-shift control} (motivated in Sec.~\ref{sec:intro}) isolates genuinely turn-aligned
timing. Within each utterance we circularly shift the prosodic features in time by a random,
nonzero offset, so each feature is realigned against a different point in the same turn. This keeps
the feature's marginal distribution and its within-utterance autocorrelation intact while
destroying its alignment with the turn boundary, and it leaves the encoder features, the labels, and
the speaker grouping untouched. We average over several random offsets and report the corrected
value as real minus time-shifted. Whatever value survives the shift cannot come from boundary
timing, and we read it as a \emph{speaker-identity shortcut}, in which a feature's roughly fixed
per-speaker level predicts boundaries through its correlation with how that speaker's turns tend to
be arranged.

\subsection{Statistical significance and splits}
\label{sec:method-ci}
We test significance with a bootstrap that resamples whole speakers rather than individual frames,
since frames within a speaker are correlated, following standard practice for confidence intervals
in speech-system evaluation \cite{bisani2004}. We draw speaker blocks with replacement $1{,}000$
times, recompute $\dbits$ and $\dauc$ on each resample, and report the $2.5$ and $97.5$ percentiles
as a $95\%$ confidence interval, treating an effect as present when the interval excludes zero and
negligible when it sits close to zero on both sides. The data is partitioned by speaker into three
disjoint splits, with about $50\%$ of the frames for fitting the detector and tuning its
regularisation, $25\%$ for any feature or operating-point selection, and $25\%$ held out for
reporting. Groups are formed by speaker and assigned so that no group spans two splits, so a
detector cannot do well by memorising a talker, and the reported numbers and intervals all come from
speakers seen in neither fitting nor selection.

\section{Experiments and Results}
\label{sec:results}

We take the three questions from the introduction in turn, then ask whether the answer generalises
beyond one encoder and whether it survives our design choices. Two guards, both defined above, run
under every measurement, and we report each one as we go. The neural detector reruns each test, so a
result can never be blamed on a detector too weak to use the prosody it is given. The time-shift
control recomputes each test with the prosodic features shifted off the turn boundary, so any value
that survives the shift is a speaker-identity shortcut rather than real timing.

\begin{table}[b]
\caption{Value prosody adds over the encoder at each depth ($W{=}10$), in bits/frame ($\dbits$)
and AUC ($\dauc$). Corrected columns remove the speaker-identity shortcut via the time-shift
control.}
\label{tab:curve}
\centering
\small
\setlength{\tabcolsep}{3.5pt}
\resizebox{\columnwidth}{!}{%
\begin{tabular}{llccccc}
\toprule
& representation & AUC($R$) & $\dbits$ & $\dbits$ corr. & $\dauc$ & $\dauc$ corr. \\
\midrule
\multirow{4}{*}{\rotatebox[origin=c]{90}{logistic}}
 & fbank      & 0.685 & $+0.053$ & $+0.080$ & $+0.092$ & $+0.091$ \\
 & zip-early  & 0.874 & $+0.021$ & $+0.016$ & $+0.014$ & $+0.007$ \\
 & zip-mid    & 0.908 & $+0.015$ & $+0.008$ & $+0.006$ & $+0.003$ \\
 & zip-final  & 0.893 & $+0.018$ & $+0.004$ & $+0.006$ & $+0.0015$ \\
\midrule
\multirow{4}{*}{\rotatebox[origin=c]{90}{neural}}
 & fbank      & 0.686 & $+0.181$ & $+0.184$ & $+0.153$ & $+0.135$ \\
 & zip-early  & 0.884 & $+0.030$ & $+0.021$ & $+0.012$ & $+0.009$ \\
 & zip-mid    & 0.912 & $+0.019$ & $+0.018$ & $+0.010$ & $+0.007$ \\
 & zip-final  & 0.901 & $+0.024$ & $+0.014$ & $+0.006$ & $+0.006$ \\
\bottomrule
\end{tabular}}
\end{table}

\subsection{How much does prosody add, and where?}
\label{sec:res-curve}
We compare the detector on the encoder alone against the same detector on the encoder plus prosody,
at each depth, in bits per frame ($\dbits$) and AUC ($\dauc$) (Table~\ref{tab:curve}). Because the
encoder alone already ranks frames well, $\dauc$ has little room to move, so we read the effect
mainly from $\dbits$.

\begin{figure}[t]
\centering
\includegraphics[width=0.96\columnwidth]{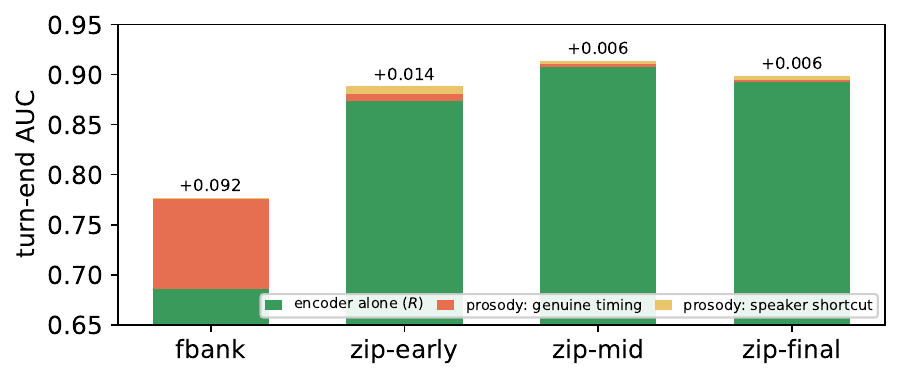}
\caption{Encoder-alone accuracy (AUC, green) and the extra accuracy from hand-built prosody, by
depth. The prosody increment splits into genuine timing (orange, removed by the time-shift
control) and a speaker-identity shortcut (yellow, which survives). Numbers above bars are the
total $\dauc$.}
\label{fig:dualbar}
\end{figure}
\begin{figure}[t]
\centering
\includegraphics[width=0.95\columnwidth]{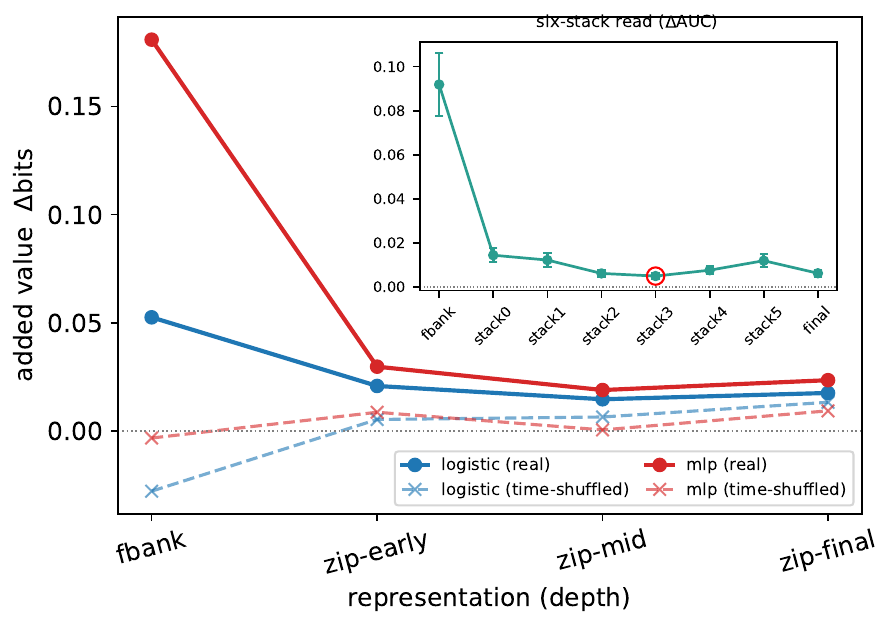}
\caption{Value prosody adds (bits/frame) by depth, with solid lines for the real features and
dashed lines for the time-shifted control, for both detectors. The inset shows a finer six-stack
read ($\dauc$, speaker-bootstrap CIs), and the circled point is the final layer.}
\label{fig:curve}
\end{figure}

\noindent\textbf{The value collapses in a single encoder stack.}
Prosody's value does not fade gradually through the encoder. It collapses in one step, right at
the boundary between the raw input and the first encoder stack (Fig.~\ref{fig:dualbar}). On the
raw filterbank features prosody adds a large $\dbits{=}{+}0.053$, but a single stack later most
of that gain is gone, down to $+0.021$, and it stays small through the rest of the encoder
(Table~\ref{tab:curve}). The neural detector shows the same collapse, so a detector too weak to
use prosody is not the explanation. A finer read across all six encoder stacks tells the same
story, the added value staying well below its raw-input level throughout (Fig.~\ref{fig:curve}).
The final layer is also the deployment-relevant one. Real turn detectors read a recogniser's
learned outputs, its encoder or joint-network states \cite{chang2022_turntaking}, transcript
\cite{livekit2024_eou}, or a pretrained speech encoder \cite{vogent2025_voturn}, not a separate
hand-built prosodic front-end. Prosody thus adds least exactly where a detector would use it.

\noindent\textbf{What is left identifies the speaker, not the turn end.}
Figure~\ref{fig:dualbar} separates the two. Each bar splits the prosody gain into the part the
time-shift control removes, which is genuine boundary timing (orange), and the part that
survives the control, which is the speaker-identity shortcut (yellow). On the raw fbank input the
bar is almost all orange, so there prosody really is reading turn timing. That orange part shrinks
away inside the encoder, and by the final layer the genuine-timing gain is only
$\dauc{=}{+}0.0015$, nothing in practice. What is left of the already tiny $\dauc{=}{+}0.006$ total
is mostly the yellow speaker-identity part. A speaker's average pitch or rate barely changes within
a turn, so it cannot mark when the turn ends, yet it still predicts boundaries because different
speakers tend to end their turns in different ways. A second control confirms this, since replacing
every frame with the utterance's own average prosody, which leaves only the per-speaker level,
recovers the same small gain, so the residue really is speaker identity. The neural detector leaves a
slightly larger residual, but at $\dauc{=}{+}0.006$ it too is negligible.


\noindent\textbf{Prosody alone and silence alone both trail the encoder.}
Two doubts could weaken the result, and a baseline settles each one. The first doubt is that
prosody might carry little turn-end information to begin with. To check this, we run a detector
on prosody alone, and it reaches AUC $0.79$. That is below every encoder depth but well above
chance, so prosody does carry real turn-end information, and the encoder has picked up almost
all of it. The second doubt is that the encoder might have learned only to spot silence rather
than the prosodic cues we care about. To check this, we run a detector on voice activity alone,
a speech flag and silence duration, and it reaches only AUC $0.63$. If we tune it to catch
$90\%$ of real turn ends, it also goes off on $99.6\%$ of the pauses in the middle of a turn,
when the speaker has not actually stopped, so it fires almost all the time. Silence on its own
cannot tell a turn end from a mid-turn pause \cite{heldner2010,raux2008}, so the encoder must
be using far more than silence to reach its accuracy.

\noindent\textbf{At an operating point, prosody barely lowers false alarms.}
The metrics so far are threshold-free, but a deployed endpointer runs at a fixed operating point,
where what matters is how often it fires before the turn has actually ended. We therefore set each
detector to $90\%$ recall of true turn ends on the collar and measure its false-alarm rate, the
share of mid-turn interior frames it wrongly flags, for $R$ and for $R{+}P$ (Table~\ref{tab:fa}).
The detection view agrees with the added-value curve. On the raw filterbank prosody cuts the
false-alarm rate by $0.11$ (from $0.71$ to $0.60$), a reduction the time-shift control confirms is
genuine turn timing. That reduction collapses inside the encoder, and at the final layer prosody
lowers false alarms by only $0.026$ ($0.356{\to}0.330$), of which the control leaves just $+0.011$
as genuine. So even at a concrete operating point, a prosodic front-end buys almost no reduction in
early firing at the deep encoder, and what little it buys is mostly the speaker-identity shortcut.
This is the within-collar operating point, not a full streaming endpointer, which must also decide
through the post-turn silence and is judged on latency (Sec.~\ref{sec:limits}).

\begin{table}[t]
\caption{False-alarm rate at $90\%$ recall on the collar (logistic, $W{=}10$), for the encoder
alone ($R$) and with prosody ($R{+}P$). $\dfa$ is the FA reduction from prosody, with a
speaker-bootstrap $95\%$ CI, and the corrected column removes the speaker-identity shortcut. A
corrected value can slightly exceed the raw $\dfa$ when time-shifted prosody raises the error
rate.}
\label{tab:fa}
\centering
\small
\setlength{\tabcolsep}{4pt}
\begin{tabular}{lcccc}
\toprule
representation & FA($R$) & FA($R{+}P$) & $\dfa$ (CI) & $\dfa$ corr. \\
\midrule
fbank      & 0.711 & 0.598 & $+0.113$\,[.084,.144] & $+0.121$ \\
zip-early  & 0.385 & 0.343 & $+0.041$\,[.024,.055] & $+0.020$ \\
zip-mid    & 0.291 & 0.271 & $+0.019$\,[.009,.029] & $+0.011$ \\
zip-final  & 0.356 & 0.330 & $+0.026$\,[.012,.039] & $+0.011$ \\
\bottomrule
\end{tabular}
\end{table}

\subsection{Why does prosody become redundant?}
\label{sec:res-recon}
Section~\ref{sec:res-curve} shows that prosody stops helping once it enters the encoder, but not
why. The natural explanation is that the encoder already holds the prosody, so a second
hand-built copy has nothing to add. We can test this directly by trying to read each prosodic
feature back out of the encoder. At every depth we predict the feature from a short causal
window of activations, the current frame and the four before it, about $200$\,ms in all, and
report the held-out $R^2$. A high $R^2$ means the encoder still represents the feature, so
supplying it again is redundant. A low $R^2$ means the encoder has thrown it away. We read out
with both a linear ridge regressor and a small neural network, so a low score cannot simply be
the limit of a linear probe.

\noindent\textbf{The encoder holds prosody most where prosody helps least.}
Reconstruction (Table~\ref{tab:recon}) runs opposite to the added-value curve
(Fig.~\ref{fig:curve}). As prosody's added value falls through the early and middle stacks
($0.053{\to}0.021{\to}0.015$ bits), its reconstructability climbs the other way ($R^2{=}0.33$ on
raw features to $0.53$ and $0.48$ inside the encoder). The two trends move against each other
because a hand-built copy adds the least exactly where the encoder already represents prosody best.
The added-value curve and the reconstruction probe therefore agree that prosody becomes redundant
once the encoder has absorbed it.

\begin{table}[t]
\caption{Median reconstruction $R^2$ of prosody from a short causal encoder window, by depth
(linear and neural reconstructors). Broadband energy is a present-signal control.}
\label{tab:recon}
\centering
\small
\setlength{\tabcolsep}{4pt}
\begin{tabular}{lccc}
\toprule
depth & prosody (lin) & prosody (MLP) & energy (control) \\
\midrule
fbank      & 0.33 & 0.40 & 1.00 \\
zip-early  & 0.53 & 0.51 & 0.98 \\
zip-mid    & 0.48 & 0.42 & 0.91 \\
zip-final  & 0.20 & 0.09 & 0.71 \\
\bottomrule
\end{tabular}
\end{table}

\noindent\textbf{At the final layer it keeps only what the task needs.}
The rise does not last, and at the final layer reconstructability drops to $R^2{=}0.20$ (ridge)
and $0.09$ (neural), so the encoder builds a prosodic representation in the middle and then strips
it to the turn-relevant part. Even at its peak the probe recovers only about half the signal, so
the encoder keeps no clean copy, only enough to make a hand-built one redundant. How it retains
this prosody is partly lexical, since deep ASR layers carry word-level information
\cite{pasad2024words}, but this cannot be the whole story, because a self-supervised encoder
trained with no word labels shows the same redundancy (wav2vec2, Sec.~\ref{sec:res-generality}).

\noindent\textbf{The final-layer drop is lost information, not a limit of the probe.}
A low final-layer $R^2$ could instead mean the probe failed, so we rule out two ways it might. Does
the probe work at that depth? Broadband energy, plainly in the audio, is recovered at every depth,
slipping only to $0.71$ at the final layer. Is prosody too rich to decode? The same probe reads it
well at the middle stacks ($0.53$, $0.48$), so difficulty is constant across depths. Widening the
window from $5$ to $11$ frames does not recover it either ($0.20$ to $0.17$), so the information is
not merely spread across frames. The final-layer drop is therefore a genuine loss of prosodic
detail, not a probe artefact.

\subsection{Which features, if any, still matter?}
\label{sec:res-subset}
Testing the whole bank at once can hide a standout, because a single feature that captures turn
timing the encoder missed could be drowned out by the rest and leave the bank looking redundant. To
catch such a case, we test each feature on its own, measuring its added value with and without the
time-shift control and reading the drop as the share that comes from genuine turn timing.

\noindent\textbf{The strongest-looking features are really the speaker shortcut.}
Before the control, the features with the most apparent value at the final layer are activity
and voice-quality cues, the speech flag and normalised silence duration. The time-shift control
removes almost all of it, which places them with the speaker-identity shortcut from
Sec.~\ref{sec:res-curve} rather than with real timing. Once corrected, what remains is a scatter
of pitch and spectral features worth about $0.001$ bits each, with no clear winner.

\begin{figure}[t]
\centering
\includegraphics[width=0.95\columnwidth]{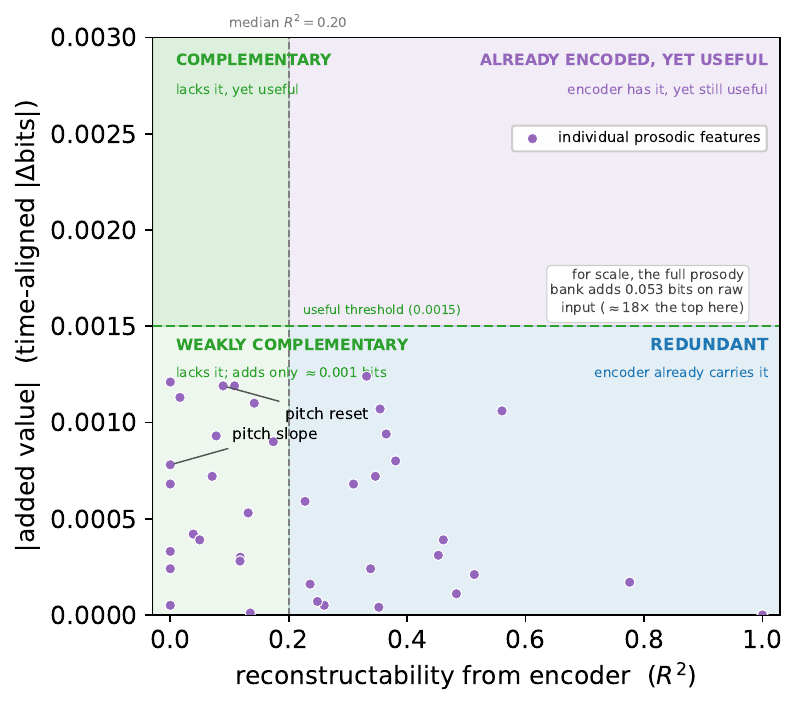}
\caption{Each final-layer prosodic feature by encoder reconstructability ($R^2$, x) and
time-aligned added value ($|\dbits|$, y). The reference lines mark the median $R^2$ ($0.20$)
and our usefulness threshold ($0.0015$).}
\label{fig:quadrant_pi}
\end{figure}

\noindent\textbf{No single feature clears our usefulness line.}
Figure~\ref{fig:quadrant_pi} shows this directly, placing each feature by how well the encoder
reconstructs it (x-axis) against how much real turn timing it adds (y-axis). A feature worth
keeping would sit high and to the left, hard for the encoder to reconstruct yet still adding real
timing, and that corner is empty. The best surviving feature adds only about $0.001$ bits,
roughly $50\times$ less than the $0.053$ bits prosody adds on raw input, so no reasonable
threshold would keep any of them, including the pitch cues that classical end-of-turn detectors
relied on most. Feature by feature, the result points the same way as the bank as a whole.

\subsection{Does this generalize to other encoders and data?}
\label{sec:res-generality}
Every result so far comes from one encoder and one set of recordings, so the pattern might hold
only for that encoder or data, so we check both.

\begin{figure}[t]
\centering
\includegraphics[width=0.95\columnwidth]{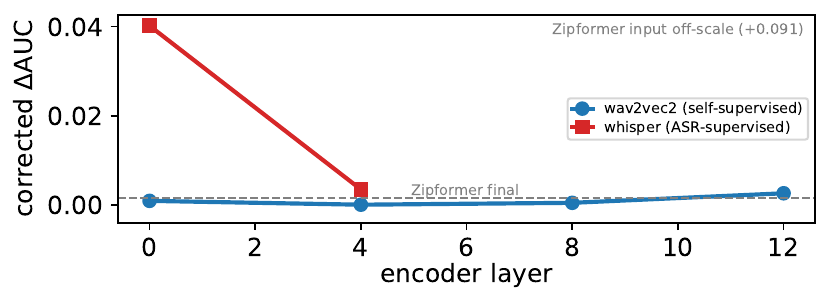}
\caption{Time-shift-corrected value prosody adds on two open-source encoders, by layer. Grey lines
mark our encoder's input and final-layer values for reference.}
\vspace{-0.2cm}
\label{fig:generality}
\vspace{-0.2cm}
\end{figure}

\noindent\textbf{The collapse is not specific to our encoder.}
We rerun the added-value measurement, with the time-shift control, on two open-source encoders.
wav2vec2 \cite{baevski2020wav2vec2} is self-supervised and never sees a recognition label, while
Whisper \cite{radford2023whisper} is trained for recognition on a different architecture. Both
expect $16$\,kHz audio, so we upsample our $8$\,kHz telephone speech and run them off the shelf,
which makes this supporting evidence rather than an exact replication. On Whisper the corrected
value starts at $+0.040$ on the input and falls by an order of magnitude to $+0.0035$ within a
few layers, the same one-step collapse as on our own encoder (Fig.~\ref{fig:generality}). On
wav2vec2 it never rises, staying below $0.003$ at every layer, because the encoder already
reaches AUC $0.95$ to $0.98$ on its own. wav2vec2 is the telling case, since it is trained with no
recognition labels and never learns the words, yet prosody is just as redundant on it, which rules
out the reading Sec.~\ref{sec:res-recon} could not exclude, that the encoder absorbs prosody
through lexical content alone.

\noindent\textbf{The result reproduces on fully public data.}
Both encoder checks still rely on our own recordings. To remove that dependence we repeat the
measurement on the public CASPER corpus \cite{casper2025} of two-party English conversation, with
the same feature bank, annotation pipeline, and time-shift control (Table~\ref{tab:casper}). Adding
a standard $80$-dim log-mel front-end gives a raw-input reference and reproduces the full
raw-to-encoder collapse, not just the across-layer redundancy. The picture matches our main study,
since raw log-mel prosody adds $\dauc{=}{+}0.085$, of which $+0.059$ survives the control as genuine
timing, whereas at any learned layer the corrected gain falls to at most $0.002$ even though the
encoders reach AUC $0.90$ to $0.995$. The speaker-identity split recurs too, with Whisper's early
layer showing a raw $\dauc{=}{+}0.046$ that the control almost removes ($+0.001$). A public corpus,
open-source encoders, and a public pipeline thus reproduce all three findings, so none depends on
our own data or model. The log-mel front-end here is a standard $80$-dimensional filterbank rather
than our encoder's exact input, so it serves as a comparable raw-input baseline rather than an
identical one.

\begin{table}[t]
\caption{Public-data replication on CASPER using Logistic detector.}
\label{tab:casper}
\centering
\small
\setlength{\tabcolsep}{5pt}
\begin{tabular}{llccc}
\toprule
representation & layer & AUC($R$) & $\dauc$ & $\dauc$ corr. \\
\midrule
raw input & log-mel & 0.763 & $+0.085$ & $+0.059$ \\
\midrule
\multirow{2}{*}{wav2vec2-base}
 & early (0)  & 0.994 & $-0.001$ & $0.000$ \\
 & deep (12)  & 0.964 & $+0.001$ & $+0.002$ \\
\midrule
\multirow{2}{*}{whisper-small}
 & early (0)  & 0.897 & $+0.046$ & $+0.001$ \\
 & deep (12)  & 0.995 & $0.000$ & $0.000$ \\
\bottomrule
\end{tabular}
\vspace{-0.3cm}
\end{table}





\subsection{Does the result survive our design choices?}
\label{sec:res-robust}
The redundancy survives the three design choices most likely to matter. Changing the label does
nothing, since narrowing or widening the collar leaves the result unchanged and even a label that
marks every post-boundary frame lets the best single feature add at most $\dauc{=}{+}0.002$. The
feature selector makes no difference either, because several standard selectors
\cite{meinshausen2010,candes2018,liu2022dcrt,bh1995,peng2005} return much the same set yet cannot
separate genuine timing from the speaker shortcut, so our feature-level verdict rests on the
time-shift control rather than on selection. Nor does the data split matter, since all our numbers
already come from held-out speakers and retraining on the selected features and a fresh speaker
group reproduces the full-bank value ($\dauc{=}{+}0.0047$ against $+0.006$ at the final layer,
Table~\ref{tab:curve}). Across all three, the redundancy holds, so it belongs to the encoder rather
than to our setup.

\section{Limitations and Future Work}
\label{sec:limits}

Our study covers one language, telephone-band audio, and one main ASR encoder. The two-encoder and
public-data checks push against this, but both open-source encoders were run off the shelf on
upsampled audio, so a wider sweep across languages, channels, and architectures is still owed. Our
$41$-feature bank spans the major turn-taking cue families but is not exhaustive, so an omitted cue
could behave differently.

The larger limitation is what we measure. The collar scores frames around the turn boundary, so we
quantify the turn-end information prosody adds there, not how a deployed endpointer behaves. A real
system decides at every frame and is judged on latency and false alarms across the whole stream,
which our collar operating point (Table~\ref{tab:fa}) only partly captures. The added-value metric
is also detector-relative, reporting what our two detectors extract rather than the information
present in principle, and using both detectors together with the time-shift control makes a large
hidden effect unlikely though not impossible. Two questions remain open, the first being whether
the redundancy extends to other prosody-dependent tasks such as emotion recognition, and the second
being the best read-out depth, since the encoder is slightly more accurate one stack earlier ($0.91$
at zip-mid against $0.89$ at the final layer), in line with middle layers holding the richest
acoustic detail
\cite{pasad2021,delafuente2024}.

\section{Conclusion}
\label{sec:conclusion}

We investigated how much turn-relevant prosody a streaming ASR encoder already represents through
three complementary analyses, and they agree. Prosody's added value is large on raw audio but falls
sharply after a single encoder stack, and a time-shift control shows that most of the residual
reflects speaker identity rather than turn timing. Reconstruction explains this pattern, because the
encoder represents prosody most strongly in its middle layers and keeps only the turn-relevant part
at the final layer, so supplying it again adds little. The same pattern appears on two further
encoders, including one trained without word labels, so it is not merely lexical.

Prosody remains a genuine turn-taking signal, reaching AUC $0.79$ on its own, but a trained encoder
has already absorbed most of its turn-relevant part. For practitioners building on a trained
streaming encoder, a separate hand-built prosodic front-end offers limited benefit for end-of-turn
detection, even at the deep layer where a detector usually attaches, and no pared-down subset
recovers it. Confirming this inside a full
streaming endpointer, with latency and false-alarm costs, remains future work.



\bibliographystyle{IEEEtran}
\bibliography{references}


\end{document}